\documentclass[aps,pre,showpacs,amsmath,amssymb,longbibliography]{revtex4-1}
\usepackage{amsmath,amssymb,latexsym,epsfig,graphicx,epsf,bm}
\usepackage{graphicx}
\usepackage{float}

\newcommand{\bR}{\bold R}

\newcommand{\br}{\bold r}

\newcommand{\be}{\begin{equation}}
\newcommand{\ee}{\end{equation}}
\newcommand{\fig}[1]{Fig.~\ref{#1}}
\newcommand{\Fig}[1]{Figure~\ref{#1}}

\newcommand{\eq}[1]{Eq.~(\ref{#1})}

\newcommand{\Tc}{{T}_{\rm conf}}

\newcommand{\Ts}{{T}_{\rm s}}

\begin{document}
	\title{Active-matter isomorphs in the size-polydisperse Ornstein-Uhlenbeck Lennard-Jones model}
	\date{\today}
        \author{Daniel Jespersen}
	\affiliation{\textit{Glass and Time}, IMFUFA, Department of Science and Environment, Roskilde University, P.O. Box 260, DK-4000         Roskilde, Denmark}
	\author{Lorenzo Costigliola}
        \affiliation{\textit{Glass and Time}, IMFUFA, Department of Science and Environment, Roskilde University, P.O. Box 260, DK-4000 Roskilde, Denmark}	
	\author{Jeppe C. Dyre}
	\affiliation{\textit{Glass and Time}, IMFUFA, Department of Science and Environment, Roskilde University, P.O. Box 260, DK-4000         Roskilde, Denmark}	
	\author{Shibu Saw}\email{shibus@ruc.dk}
	\affiliation{\textit{Glass and Time}, IMFUFA, Department of Science and Environment, Roskilde University, P.O. Box 260, DK-4000         Roskilde, Denmark}
 
\begin{abstract}
    This paper studies size-polydisperse Lennard-Jones systems described by active Ornstein-Uhlenbeck particle dynamics. The focus is on the existence of isomorphs (curves of invariant structure and dynamics) in the model's three-dimensional phase diagram. Isomorphs are traced out from a single steady-state configuration by means of the configurational-temperature method. Good invariance of the reduced-unit radial distribution function and the mean-square displacement as a function of time is demonstrated for three uniform-distribution polydispersities, $12\%$, 23\%, and 29\%. Comparing to active-matter isomorphs generated by the analytical direct-isomorph-check method, the latter give somewhat poorer invariance of the structure, but better invariance of the dynamics. We conclude that both methods can be used to quickly get an overview of the phase diagram of polydisperse AOUP models involving a potential-energy function obeying the hidden-scale-invariance property required for isomorph theory to apply. 
\end{abstract}

\maketitle

\section{Introduction}\label{Intro}

Active matter involves particles that absorb energy from their environment and continuously perform motion dissipated into heat. This kind of motion, which in contrast to standard Newtonian or Brownian dynamics breaks time-reversal invariance \cite{man20a,byr22}, is relevant not only for describing biological systems ranging from bacteria to flocking birds \cite{ang11a,mar13a,bec16,ram17,sai18,das20,sha21,bow22}, but also for microscopic artificial microswimmers and active Janus particles. 

Many different approaches to the description of active matter exist, depending on whether point particles or particles with directional coordinates are considered and depending on the precise mechanism by which the particles autonomously perform mechanical work \cite{mar13,tak15,bec16,ram17,das20,sha21}. Point-particle active-matter models include the Active Brownian Particle (ABP) \cite{klo19,but22} and Active Ornstein-Uhlenbeck Particle (AOUP) models; these models have been used to describe the motion, e.g., in active colloids \cite{mag14}. The AOUP model, which is simpler than the ABP model and has one less parameter, can be used to approximate ABP dynamics. Moreover, the AOUP model offers more possibilities to obtain theoretical predictions \cite{far15,cap22}; this is the model we choose to study in the present paper. Specifically, the AOUP model involves point particles subject to a colored-noise Langevin dynamics \cite{far15,mag15,sza15,fod16}. 

In view of the variability of biological and other active systems, one cannot expect all particles to be identical. As a consequence, polydispersity has recently come into focus in connection with active-matter models \cite{ni15,hen20,kum21,sza21}. There is also currently great deal of interest in passive polydisperse systems coming from, in particular, their use in SWAP-equilibrated supercooled liquids \cite{nin17}, in which context the question arises of how similar the dynamics of small and large particles are \cite{abr08b,zac15,pih23}. Finally, it is worth mentioning that active matter at high density has recently been studied inspired by biological materials such as cells, both for monodisperse \cite{cap20a,sza21} and polydisperse cases \cite{ket22}, showing emerging collective phenomena with the spontaneous occurrence of spatial velocity correlations.

This paper studies the size-polydisperse AOUP Lennard-Jones (LJ) model. We recently demonstrated the existence of lines of approximately invariant structure and dynamics in the phase diagram of a binary LJ AOUP model; such lines are referred to as ``active-matter isomorphs'' \cite{IV,saw23a,saw23b}. Inspired by the fact that the introduction of polydispersity into ordinary (passive) Newtonian models does not affect the existence of isomorphs \cite{ing21}, the present paper investigates whether the existence of isomorphs also survives the introduction of polydispersity into the AOUP model. This is worthwhile to investigate since the existence of isomorphs makes it possible to quickly establish an overview of the phase diagram because only a single point on each isomorph needs to be simulated.

\section{The AOUP equation of motion and simulation details}\label{sim}

We consider a system of $N$ particles in volume $V$ and define the number density by $\rho\equiv N/V$. If the potential-energy function is denoted by $U(\bR)$ in which $\bR\equiv (\br_1,...,\br_N)$ is the vector of all particle coordinates, the AOUP equation of motion \cite{far15,mag15,sza15,fod16} is

\be\label{EOM_AOU}
\dot{\bR}
\,=\, -\mu \nabla U(\bR)\,+\,\bm\eta(t)\,.
\ee
Here $\mu$ is the mobility (velocity over force); the noise vector $\bm\eta(t)$ is colored according to an Ornstein-Uhlenbeck process, i.e., is a Gaussian stochastic process characterized by 

\be\label{noise}
\langle \eta_i^\alpha(t)\eta_j^\beta(t')\rangle
\,=\,\delta_{ij}\delta_{\alpha\beta}\frac{D}{\tau}\,e^{-|t-t'|/\tau}\,
\ee
in which $i$ and $j$ are particle indices, $\alpha$ and $\beta$ are $xyz$ spatial indices, and $D$ and $\tau$ are constants, respectively, of dimension length squared over time and time. 

We are interested in how the physics is affected when the density is changed, specifically in determining whether approximately invariant physics can be obtained by adjusting $D$ and $\tau$ properly with density ($\mu$ is regarded as a material constant throughout). For the binary AOUP model this problem was studied in Ref. \onlinecite{saw23a} that demonstrated how to change $D$ and $\tau$ with density in order to achieve invariant structure and dynamics to a good approximation. The question is whether this is possible also for systems with large size polydispersity. In the AOUP model ``reduced'' quantities are defined by using $l_0=\rho^{-1/3}$ as the length unit and $t_0=\tau$ as the time unit \cite{saw23a}. Reduced quantities are marked by a tilde. When we speak about approximately invariant structure and dynamics, it refers to this particular state-point-dependent unit system.

We studied a system of $N=5000$ particles in three dimensions interacting by Lennard-Jones (LJ)  pair potentials, which between particles $i$ and $j$ are given by $v_{ij}(r)=4\varepsilon\left[(r/\sigma_{ij})^{-12}-(r/\sigma_{ij})^{-6}\right]$ with $\sigma_{ij}=(\sigma_{i} + \sigma_{j})/2$ (Lorentz-Berthelot mixing rule) and $\varepsilon=1.0$. The particles sizes $\sigma_i$ are distributed according to a uniform distribution with unity average. As usual, the polydispersity $\delta$ is defined by $\delta^2=(\langle \sigma^2\rangle-\langle \sigma\rangle^2)/\langle \sigma\rangle^2$, which in our case reduces to $\delta^2=\langle \sigma^2\rangle-1$. For a uniform distribution $\delta$ cannot exceed $1/\sqrt{3}\cong 58$\%. The three polydispersities studied below are $\delta\cong 11.5\%$, 23.1\%, and 28.9\%, corresponding to the size ranges listed in Table \ref{polytable} (for brevity these  are henceforth reported as $\delta=12\%$, 23\%, and 29\%). Note that the study entails substantially different particle sizes, with the ratio of largest to smallest particle volume equal to 27 in the 29\% polydispersity case.

\begin{table}[H]
  \begin{center}
      \begin{tabular}{|c|c|c|}\hline
              $\delta$   &      $\sigma$ range  &   $\sigma_{max}/\sigma_{min}$   \\ \hline
              $12\%$     &      $0.80-1.20$     &     $1.50$       \\ \hline
              $23\%$     &      $0.60-1.40$     &     $2.33$        \\ \hline
              $29\%$     &      $0.50-1.50$     &     $3.00$     \\ \hline
     \end{tabular}
   \caption{Values of the polydispersity $\delta$, $\sigma$ range, and ratio between largest and smallest particle sizes for the three cases of uniform polydispersity studied.}
   \label{polytable}
  \end{center}
\end{table}

All simulations used a shifted-force cutoff \cite{tox11a} of the $ij$ particle interaction at the pair distance $r=2.5\sigma_{i j}$ and the time step $\Delta t = \Delta \tilde t / (D ~ \rho^{2/3})$ in which $\Delta \tilde t = 0.4$ \cite{saw23a}. The active-matter simulations were carried out on GPU cards using a home-made code, the MD simulations used RUMD \cite{RUMD}.

\section{Structure and dynamics along an isochore}

\begin{figure}[H]
\begin{center}
    \includegraphics[width=0.98\textwidth, angle=0]{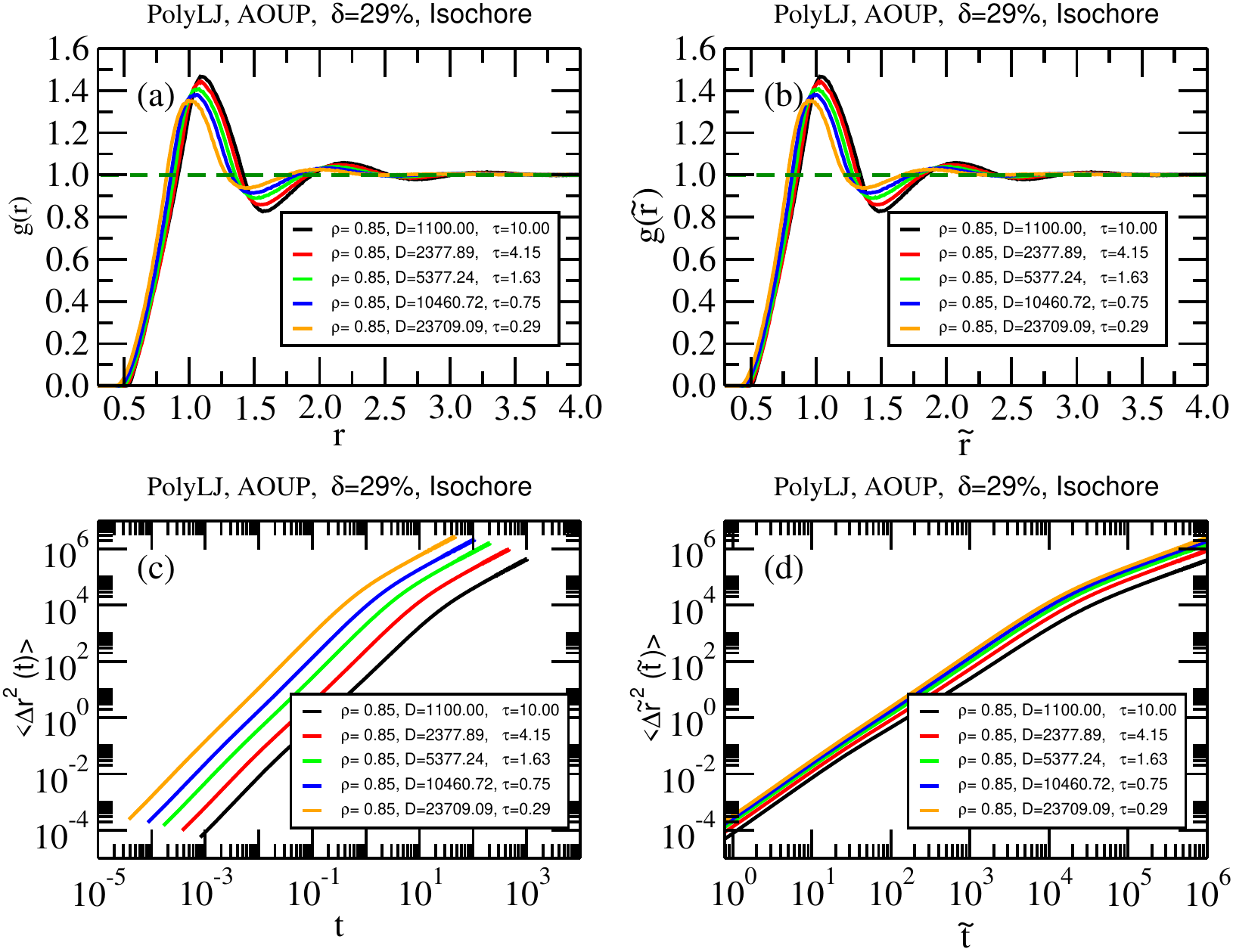}\hspace{4mm}
\caption{Average radial distribution function (RDF) and mean-square displacement (MSD) for state points on the $\rho=0.85$ isochore of the $\delta=29\%$ polydispersity LJ AOUP model (the $D$ and $\tau$ values are those of the below studied $\delta=29\%$ active-matter DIC isomorph). 
(a) and (b) show the RDF as a function of $r$ and of the reduced pair distance $\tilde{r}$, respectively (the curves are the same because $\tilde{r}\propto r$ along an isochore). We see a substantial variation in the structure, with the most pronounced structure found for the smallest values of the model parameter $D$ (black curves). The MSD likewise shows no collapse along the isochore, whether plotted (c) as a function of the time $t$ or (d) as a function of the reduced time $\tilde{t}$. The slowest motion is found for the smallest $D$ (black curves).}
	\label{fig1}
	\end{center}
\end{figure}

Before discussing results for the variation of structure and dynamics along active-matter isomorphs, we briefly present analogous results along an isochore, i.e., for state points of the same density. This sets the stage by illustrating that structure and dynamics do vary significantly throughout the $(\rho,D,\tau)$ AOUP phase diagram. Structure is studied by means of the average radial distribution function (RDF) denoted by $g(r)$. In \fig{fig1}(a) RDFs are shown along the $\rho=0.85$ isochore for the $\delta=29$\% case, with values of $D$ and $\tau$ taken from the $\delta=29$\% DIC active-matter isomorph studied below. \Fig{fig1}(b) shows the same data in reduced coordinates, which in this case simply involves in a common scaling of the x-coordinate. The parameters used in the simulations are listed in insets of the figures (more decimals of these parameters are provided in the Appendix). 

We find a substantial structure variation along the isochore. The same applies for the mean-square displacement (MSD) as a function of the time $t$, $\langle\Delta r^2(t)\rangle$, which is plotted in a log-log plot in (c) LJ units and (d) reduced units. The short-time slope is two, reflecting the ``ballistic'' regime of the AOUP model, which is not present in ordinary Langevin dynamics \cite{far15,mag15,sza15,fod16} because it results from short-time noise correlations resulting in an inertia-like persistence of the direction of motion. At long times the well-known diffusive behavior leading to unity slope is observed. We note that the dynamics varies significantly along the isochore, whether or not reported in reduced units.

\section{Structure and dynamics along $\Tc$-generated active-matter isomorphs}

Reference \onlinecite{saw23a} used the \textit{configurational temperature} $\Tc$ for determining how to change the AOUP model parameters $D$ and $\tau$ with density in order to achieve (approximately) invariant reduced structure and dynamics. The assumption is that $k_B\Tc$ is the relevant characteristic energy scale where $\Tc$ is defined by $k_B\Tc\equiv\langle(\nabla U)^2\rangle/\langle\nabla^2 U\rangle$ \cite{LLstat,rug97,pow05} in which $k_B$ is the Boltzmann constant, $\nabla$ is the gradient operator in the $3N$-dimensional configuration space, and the sharp brackets denote standard canonical-ensemble averages. In the thermodynamic limit the relative fluctuations of both the numerator and the denominator of $\Tc$ go to zero, which implies that it is enough to consider a single configuration $\bR_0$ using the expression $k_B\Tc\cong (\nabla U(\bR_0))^2/\nabla^2 U(\bR_0)$.

The reasoning of Ref. \onlinecite{saw23a} may be summarized as follows. Adopting $e_0=k_B\Tc$ as the energy unit supplementing the above introduced length and time units ($l_0=\rho^{-1/3}$; $t_0=\tau$), we first note that the three quantities $\mu t_0e_0/l_0^2$, $Dt_0/l_0^2$, and $\tau/t_0$ are dimensionless. Assuming that these quantities cannot change with varying density if the structure and dynamics are invariant in reduced units, we conclude that $\mu\propto l_0^2/(t_0e_0)=\rho^{-2/3}/(\tau k_B\Tc)$ and $D\propto l_0^2/t_0=\rho^{-2/3}/\tau$. Since $\mu$ is assumed to be a material constant, this leads to $\tau\propto\rho^{-2/3}/k_B\Tc$ and $D\propto k_B\Tc$, i.e., to the following recipe for how $D$ and $\tau$ changes with density in terms of their values $D_0$ and $\tau_0$ at a reference state point of density $\rho_0$:

\begin{eqnarray}\label{OUP_param}
D(\rho)&\,=\,&D_0\,\,\frac{\Tc(\rho)}{\Tc(\rho_0)}\,,\nonumber\\
\tau(\rho)&\,=\,&\tau_0\left(\frac{\rho_0}{\rho}\right)^{2/3}\frac{\Tc(\rho_0)}{\Tc(\rho)}\,.
\end{eqnarray}
For a large system $\Tc(\rho_0)$ may be evaluated from a single (steady-state) configuration, $\Tc(\rho_0)\cong\Tc(\bR_0)$. Reference \onlinecite{saw23a} demonstrated that this approximation introduces a negligible error for typical system sizes. In order to find $\Tc(\rho)$ one scales $\bR_0$ uniformly to the density $\rho$, i.e., substitutes $\bR=(\rho_0/\rho)^{1/3}\bR_0$ into the configurational temperature expression. This leads to 

\begin{eqnarray}\label{OUP_param2}
	D(\rho)&\,=\,&D_0\,\,\frac{\Tc\left[(\rho_0/\rho)^{1/3}\bR_0\right]}{\Tc(\bR_0)}\nonumber\,,\\
	\tau(\rho)&\,=\,&\tau_0\left(\frac{\rho_0}{\rho}\right)^{2/3}\frac{\Tc(\bR_0)}{\Tc\left((\rho_0/\rho)^{1/3}\bR_0\right)}\,.
\end{eqnarray}

We used these equations for generating three active-matter isomorphs starting in each case from the parameter values $D=1100$ and $\tau=10$ at the reference densities $\rho_0=$0.99, 0.91, and 0.85, respectively, for the polydispersities $\delta=12\%$, 23\%, and 29\% (the reference densities were chosen to have the same virial, i.e., give the same contributions to the pressure coming from the interactions).

\begin{figure}[H]
\begin{center}
    \includegraphics[width=0.98\textwidth, angle=0]{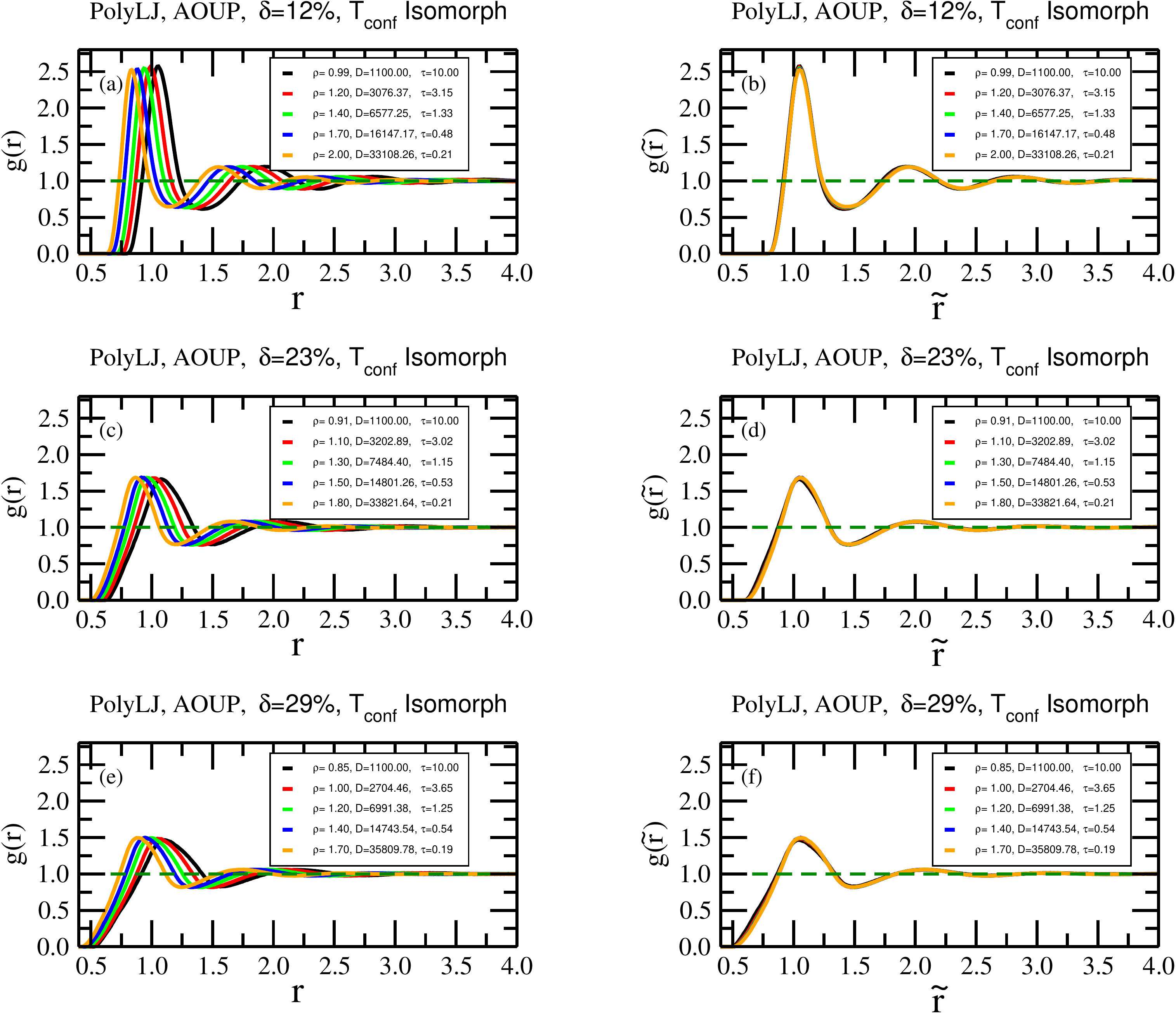}\hspace{4mm}
 \caption{Structure probed along $\Tc$-generated active-matter isomorphs. 
 (a), (c), and (e) show the average RDFs for polydispersity $\delta=12\%$, 23\%, and 29\%, respectively, while (b), (d), and (f) show the same data as functions of the reduced pair distance $\tilde{r}$. In all three cases we see a good collapse of the reduced RDF along the active-matter isomorph.}
	\label{fig2}
	\end{center}
\end{figure}

Results for the variation of the average RDF are given in \fig{fig2}. The left column reports the RDF for the three polydispersities as functions of the pair distance $r$, the right column shows the same data as functions of the reduced pair distance $\tilde{r}$. In the latter case we find a good, but not perfect, data collapse and conclude that the average structure is approximately invariant along the active-matter isomorphs. In view of the fact that the density varies by no less than a factor of two, this is not trivial.

\begin{figure}[H]
\begin{center}
        \includegraphics[width=0.98\textwidth, angle=0]{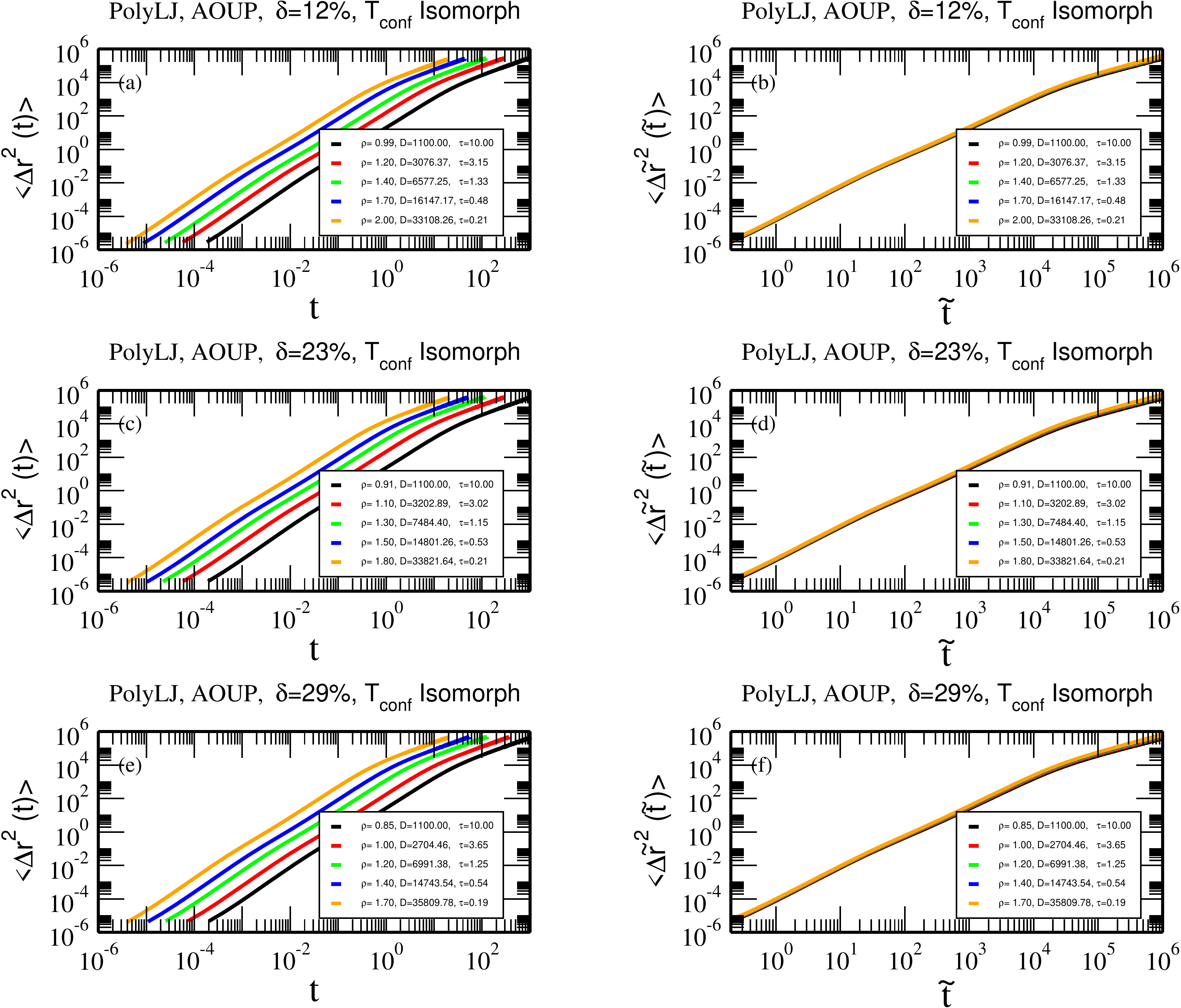}\hspace{4mm}
	\caption{Dynamics probed along $\Tc$-generated active-matter isomorphs.  
 (a), (c), and (e) show the MSDs for polydispersity $\delta=12\%$, 23\%, and 29\%, respectively, as functions of time, while (b), (d), and (f) show the same data in reduced units. There is a good, but not perfect, collapse of the reduced MSD along the active-matter isomorphs.}
	\label{fig3}
	\end{center}
\end{figure}

\Fig{fig3} shows analogous data for the MSD plotted in the same way with the left column giving the MSD as a function of time and the right column giving the same data in reduced units. There is a good data collapse with, however, a somewhat faster motion at the higher densities.

\section{comparing to direct-isomorph-check generated isomorphs}

Above we demonstrated good invariance of the structure and dynamics along active-matter isomorphs generated by the $\Tc$ method  \cite{saw23a}. That method is easy to use and efficient because it requires just a single steady-state configuration at the reference state point in order to trace out the corresponding active-matter isomorph in the relevant phase diagram, \textit{in casu} the $(\rho,D,\tau)$ diagram of the AOUP model. An alternative method for tracing out active-matter isomorphs is the analytical ``direct isomorph-check'' (DIC) method, which in Appendix A of Ref. \onlinecite{saw23a} was shown to result in somewhat better isomorph-invariance of the dynamics for the AOUP Kob-Andersen binary LJ model. 

Consider first a standard passive Newtonian systems involving LJ pair interactions of any kind, i.e., single-component, binary, or polydisperse systems, defined by some mixing rule. For such a system the analytical DIC recipe for tracing out a standard equilibrium isomorph \cite{boh12,ing12a} is 

\be\label{DIC}
\frac{h(\rho)}{T}
\,=\,{\rm Const.}
\ee
Here $h(\rho)$ is the following function of density \cite{boh12,ing12a}

\be\label{hrho}
h(\rho)
\,=\,\left(\frac{\gamma_0}{2}-1\right)\left(\frac{\rho}{\rho_0}\right)^4-\left(\frac{\gamma_0}{2}-2\right)\left(\frac{\rho}{\rho_0}\right)^2
\ee
in which $\rho_0$ is the reference-state-point density and $\gamma_0$ is the density-scaling exponent at the reference state point. The latter quantity can be determined numerically by means of 

\be\label{gamma}
\gamma_0
\,=\,\frac{\langle\Delta U \Delta W \rangle}{\langle(\Delta U)^2\rangle}\,.
\ee
in which $\Delta W$ and $\Delta U$ are the deviations from the equilibrium values of virial and potential energy, respectively, and angular brackets denote $NVT$ equilibrium averages \cite{I,IV}.

The systemic temperature $\Ts(\bR)$ is defined as the temperature of the corresponding thermal-equilibrium Newtonian system at the state point with the density of the configuration $\bR$ and average potential energy equal to $U(\bR)$ \cite{dyr20}. In the thermodynamic limit of any system (passive or active, equilibrium or non-equilibrium) fluctuations in $\Ts(\bR)$ go to zero, implying that one has at any time a well-defined systemic temperature $\Ts$. For, e.g., a driven passive or an active-matter system, a ``systemic isomorph'' is defined as a curve in the $(\rho,\Ts)$ phase diagram identical to an isomorph in the standard equilibrium Newtonian $(\rho,T)$ phase diagram \cite{dyr20}. Thus in the analytical DIC, the systemic-temperature's variation with density is given by

\be\label{DIC_Ts}
\frac{h(\rho)}{\Ts(\rho)}
\,=\,{\rm Const.}\,,
\ee
i.e., $\Ts(\rho)\propto h(\rho)$. The analytical DIC method for generating an active-matter isomorph of the AOUP model is arrived at by replacing the configurational temperature in \eq{OUP_param} by the systemic temperature $\Ts$ (this procedure is justified in Ref. \onlinecite{saw23a}). Via \eq{DIC_Ts} this leads to 

\begin{eqnarray}\label{OUP_param_DIC}
D(\rho)&\,=\,&D_0\,\,\frac{h(\rho)}{h(\rho_0)}\,,\nonumber\\
\tau(\rho)&\,=\,&\tau_0\left(\frac{\rho_0}{\rho}\right)^{2/3}\frac{h(\rho_0)}{h(\rho)}\,.
\end{eqnarray}

Table II reports the systemic temperatures at the reference state points of the three polydispersities studied. As mentioned, the reference densities were chosen to have the same virial; we see that they also have almost the same systemic temperature.

\begin{table}[H]
  \begin{center}
      \begin{tabular}{|c|c|c|c|}\hline
              $\rho_0$    &     $\delta$  &       $T_s$   &       $\langle U\rangle$        \\ \hline
              $0.990$   &     $12\%$    &       $0.96$  &       $-4.455$    \\ \hline
              $0.905$   &     $23\%$    &       $0.98$  &       $-4.447$    \\ \hline
              $0.850$   &     $29\%$    &       $1.00$  &       $-4.321$    \\ \hline
     \end{tabular}
   \caption{Systemic temperature $\Ts$ and average potential energy $\langle U\rangle$ at the reference densities $\rho_0$ of the three polydisperse systems studied. In all three cases the values of the AOUP parameters at the reference densities are $D=1100$ and $\tau=10$.}
   \label{Tstable}
  \end{center}
\end{table}

\begin{figure}[H]
	\begin{center}
    \includegraphics[width=0.42\textwidth, angle=0]{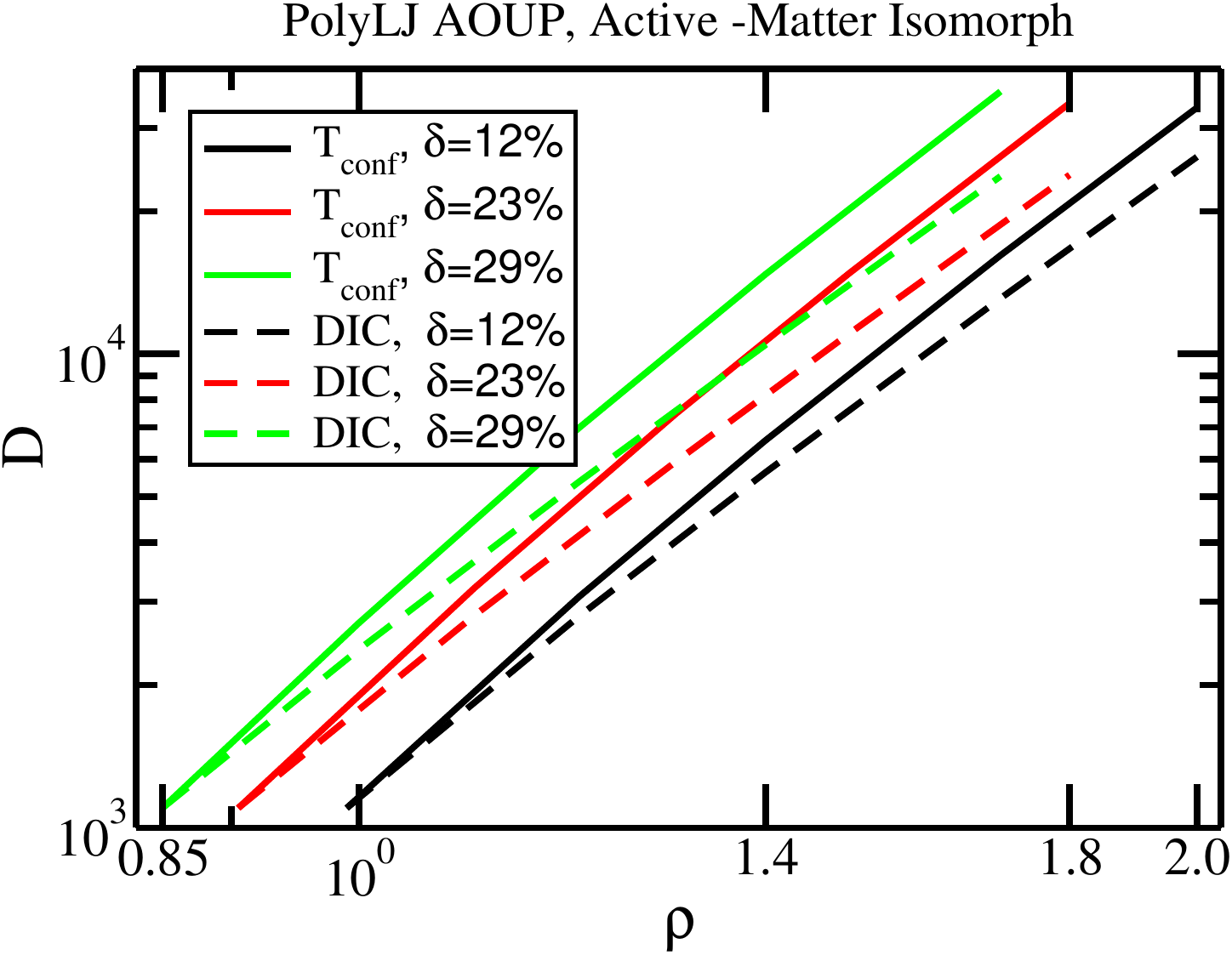}
	\caption{Active-matter isomorphs for the polydispersities $\delta=12\%$, 23\%, and 29\%, generated from the reference state points by two different methods, the $\Tc$ method of Sec. III (full curves) and the analytical direct-isomorph-check (DIC) method (dashed curves). The isomorphs are visibly different.}
	\label{fig4}
	\end{center}
\end{figure}

\begin{figure}[H]
	\begin{center}
            \includegraphics[width=0.98\textwidth, angle=0]{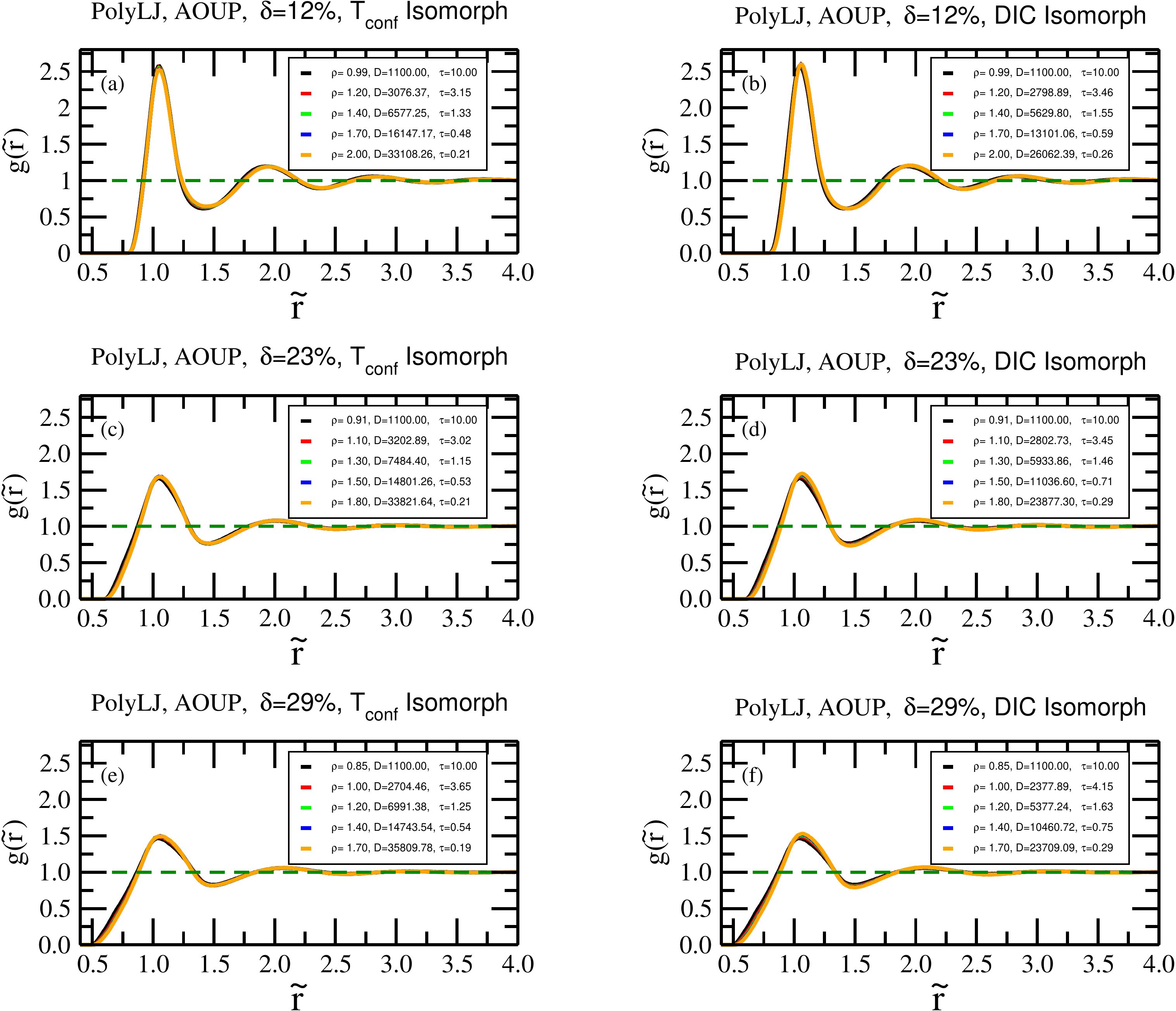}\hspace{4mm}
	\caption{Comparing the degree of structural invariance along $\Tc$-generated and DIC-generated active-matter isomorphs.
  (a), (c), and (e) show the reduced average RDFs for polydispersity $\delta=12\%$, 23\%, and 29\%, along the $\Tc$-generated isomorphs, while (b), (d), and (f) show the corresponding reduced average RDFs along the DIC-generated isomorphs. There is a somewhat better data collapse along the $\Tc$-generated isomorphs. }
	\label{fig5}
	\end{center}
\end{figure}

Fig.~\ref{fig4} shows the active-matter isomorph obtained from the $\Tc$ method (full curves) and the analytical DIC method (dashed curves), starting at the reference state point $(\rho,D,\tau)=(\rho_0,1100,10)$ in which the reference density is 0.99, 0.91, and 0.85, respectively, for $\delta=12\%$, 23\%, and 29\%. The two methods for generating isomorphs result in visibly different curves; thus there is more than 50\% difference in $D$ and $\tau$ at the largest density in the 29\% polydispersity case (green curves). How different are these active-matter isomorphs when it comes to average RDF and MSD data collapse? The RDF case is investigated in \fig{fig5}, which shows that the structure is somewhat more invariant along the $\Tc$-generated active-matter isomorphs than along the DIC-generated isomorphs, albeit this is a minor effect because in both cases the structure is fairly invariant. The differences are most pronounced at higher polydispersity.

\begin{figure}[H]
	\begin{center}
                \includegraphics[width=0.98\textwidth, angle=0]{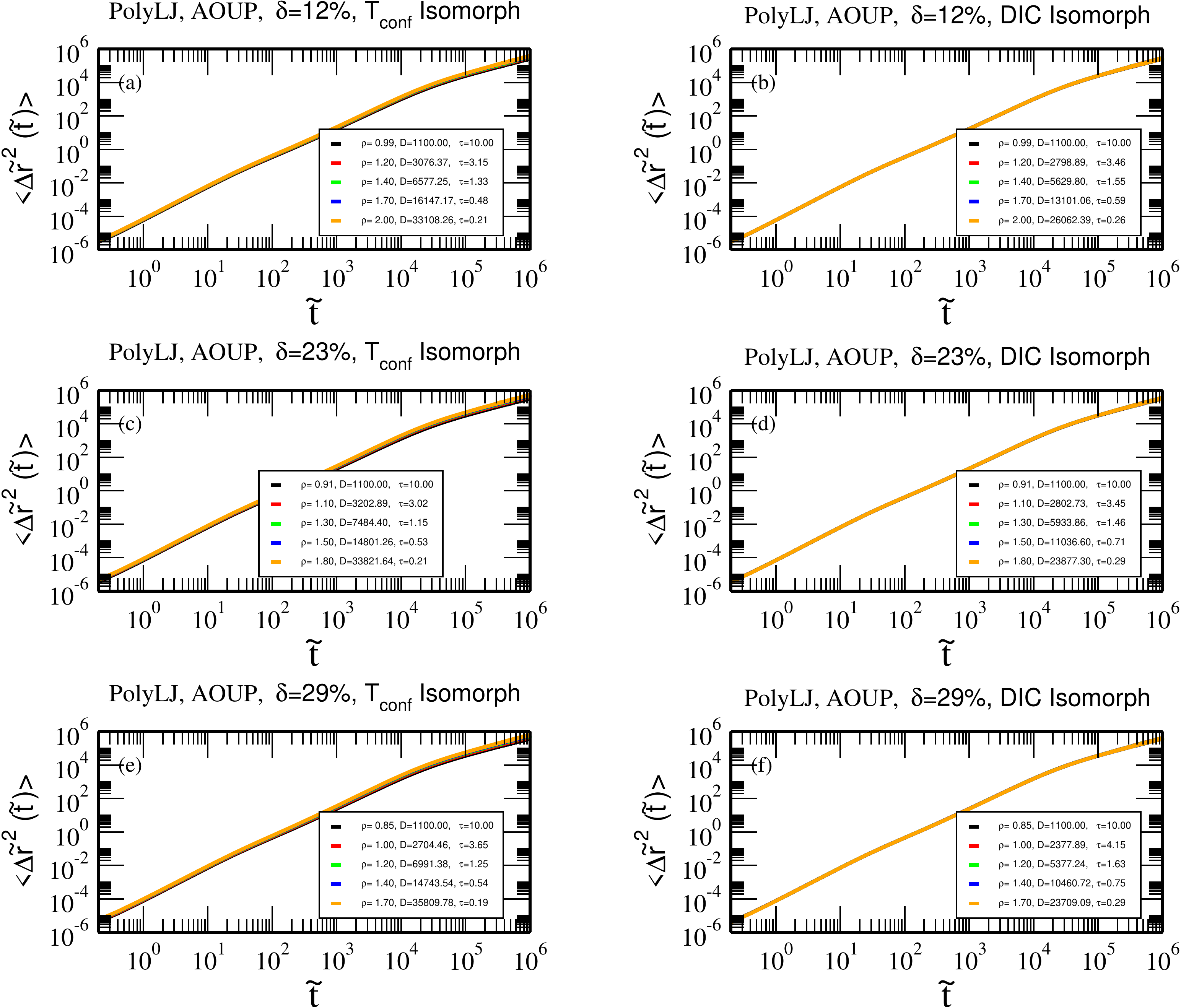}\hspace{4mm}
	\caption{Comparing the degree of invariance of the dynamics along the $\Tc$-generated and DIC-generated active-matter isomorphs.
  (a), (c), and (e) show the reduced MSDs for polydispersity $\delta=12\%$, 23\%, and 29\%, along the $\Tc$-generated isomorphs, while (b), (d), and (f) show the corresponding reduced MSDs along the DIC-generated isomorphs. There is a better data collapse along the DIC-generated isomorphs. }
	\label{fig6}
	\end{center}
\end{figure}

\Fig{fig6} reports results for the MSD. Here we reach the opposite conclusion: the DIC method results in a somewhat better data collapse than the $\Tc$ method. The same conclusion was reached for the binary Kob-Andersen AOUP model in Ref. \onlinecite{saw23a} (that did not investigate the average RDF).

\section{Role of smallest and largest particles}

\begin{figure}[H]
	\begin{center}
        \includegraphics[width=0.98\textwidth, angle=0]{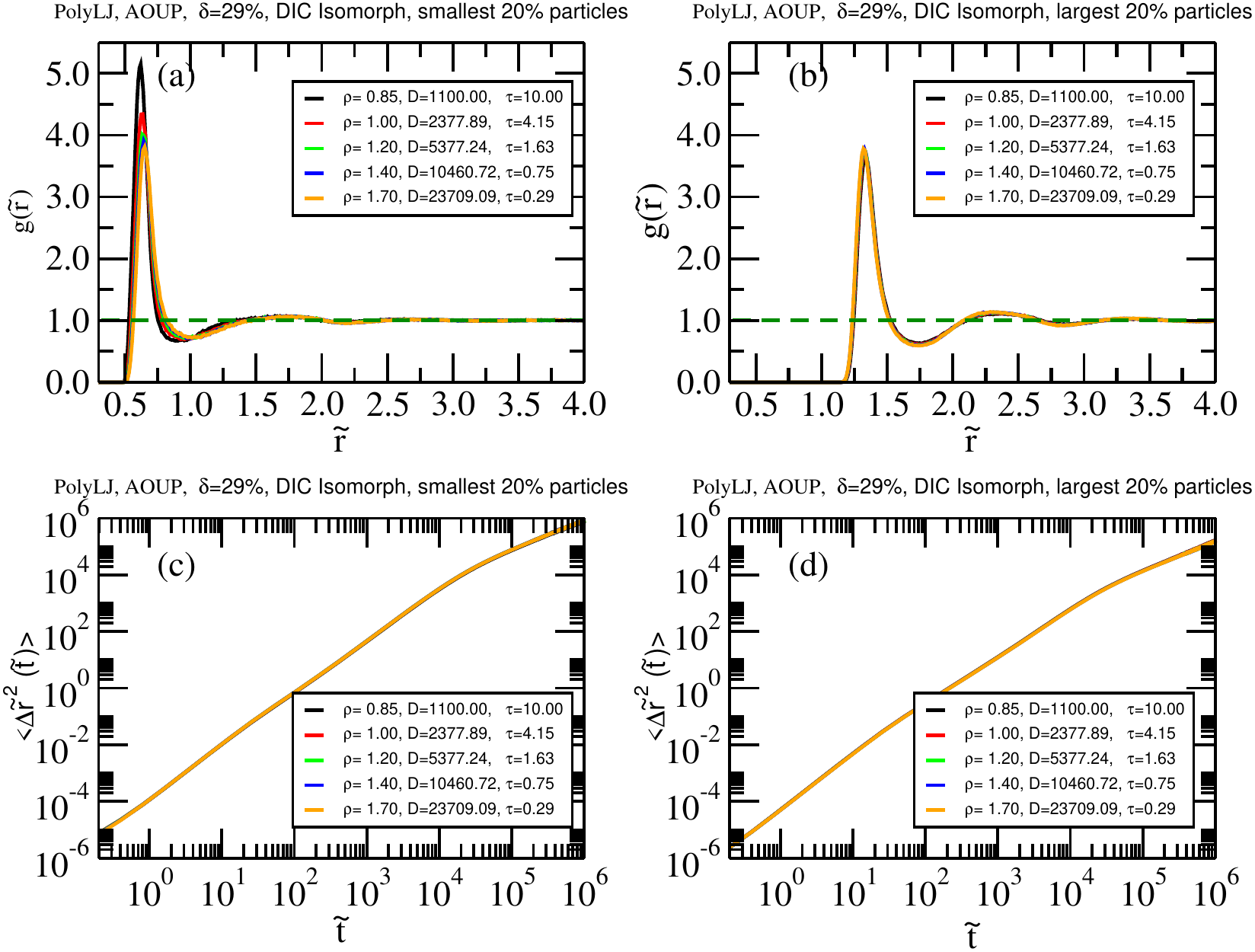}\vspace{4mm}	
	\caption{Comparing the degree of invariance of the structure and dynamics of the $20\%$ smallest and largest particles, along the DIC-generated active-matter isomorphs for the $\delta=29\%$ polydispersity case. (a) and (c) show the reduced RDFs. The invariance of the largest $20\%$ particle RDF is much better than the small-particle RDF. For the MSD, however, both cases are isomorph invariant to a good approximation.}
	\label{fig7}
	\end{center}
\end{figure}

To illuminate why the structure is not always isomorph invariant, we studied for the 29\% polydispersity simulation data the structure and dynamics of the 20\% smallest and largest particles along the DIC isomorph (\fig{fig7}). The RDF is here defined by limiting the central particle to be either among the smallest or among the largest 20\% and counting only neighboring particles of the same kind. In both cases, the peaks are narrower and higher than that of the average RDF (\fig{fig5}(f)), which reflects the limitation to similar-sized particles. We note that unlike in  \fig{fig5}(f), in \fig{fig7}(a) the most pronounced structure is seen at the lowest density. Interestingly, the structure around the smallest particles is not DIC-isomorph invariant, while that around the largest particles is; a similar result applies for the $\Tc$-generated isomorph (data not shown). In contrast to the findings for structure, the dynamics of both small and large particles is DIC-isomorph invariant to a good approximation, even though smallest particles move considerably faster than the largest ones. We conclude that the lack of perfect isomorph invariance of the overall RDF largely reflects the fact that the structure around the smallest particles is not isomorph invariant.

\section{Summary and outlook}

We have shown that the uniform-distribution size-polydisperse LJ AOUP model has active-matter isomorphs for the polydispersities $\delta=12\%$, 23\%, and 29\%. This demonstrates the robustness of the active-matter-isomorph concept, which for passive systems applies whenever the potential-energy function obeys the hidden-scale-invariance condition discussed in Refs. \onlinecite{sch14,dyr18a}. The existence of isomorphs means that the dimension of the polydisperse AOUP phase diagram is effectively reduced from three to two, since it implies that lines exist in the $(\rho,D,\tau)$ phase diagram along which the reduced structure and dynamics are invariant to a good approximation. From a practical perspective, this fact makes it easy to quickly get an overview of the AOUP model's phase diagram.

Two methods have been studied for generating active-matter isomorphs, one based on the configurational temperature and one based on the systemic-temperature concept. We find that both methods work well despite the fact that they do not trace out identical active-matter isomorphs (\fig{fig4}). In practice, the latter method will be easier to use in the case of LJ active matter for which a simple expression is available for the function $h(\rho)$ where the parameter $\gamma_0$ may be evaluated from a single passive-matter simulation (\eq{gamma}).

More work is needed to clarify how polydispersity relates to the existence of active-matter isomorphs in general. As regards the AOUP model, it would be interesting to investigate whether the introduction of energy polydispersity affects the existence of isomorphs. More generally, other models like the active Brownian particle model with a potential-energy function that obeys hidden scale invariance should be investigated in polydisperse versions in order to illuminate the robustness of the active-matter-isomorph concept.

\begin{acknowledgments}
	This work was supported by the VILLUM Foundation's \textit{Matter} grant (16515).
\end{acknowledgments}

%

\setcounter{table}{0}
\renewcommand{\thetable}{S\arabic{table}}

\setcounter{section}{0}  
\renewcommand{\thesection}{S\arabic{section}}

\clearpage
\begin{center}
    \Large\bf{Appendix}
\end{center}

This Appendix provides more decimals than given in the figures of the parameters $\rho$, $D$, and $\tau$ used in the simulations. It follows from \eq{OUP_param2} and \eq{OUP_param_DIC} that both methods of tracing out active-matter isomorphs result in $D\tau\rho^{2/3}=$Const. Within the number of decimals given, this applies for the below reported parameters. 

\begin{table}[H]
  \begin{center}
      \begin{tabular}{|l|c|r|}\hline
              $\rho$ & $D$           &     $\tau$   \\ \hline
              $0.8500 $ & $1100.0000$   &     $10.0000$ \\ \hline
              $0.8500 $ & $2377.8887$   &     $4.1509$  \\ \hline
              $0.8500 $ & $5377.2401$   &     $1.6255$  \\ \hline
              $0.8500 $ & $10460.7186$  &     $0.7540$  \\ \hline
              $0.8500 $ & $23709.0934$  &     $0.2923$  \\ \hline
     \end{tabular}
   \caption{Values of $D$ and $\tau$ for polydispersity $\delta=28.87\%$ along the $\rho=0.85$ isochore (Fig.~1).}
   \label{tableRho0.85}
  \end{center}
\end{table}

\begin{table}[H]
  \begin{center}
      \begin{tabular}{|l|c|r|}\hline
              $\rho$        &    $D$           &    $\tau$     \\ \hline
              $0.9900$      &    $1100.0000$   &    $10.0000$  \\ \hline
              $1.2000$      &    $3076.3710$   &    $3.1453$   \\ \hline
              $1.4000$      &    $6577.2502$   &    $1.3275$   \\ \hline
              $1.7000$      &    $16147.1699$  &    $0.4751$   \\ \hline
              $2.0000$      &    $33108.2613$  &    $0.2079$   \\ \hline
     \end{tabular}
   \caption{Values of $\rho$, $D$ and $\tau$ for polydispersity $\delta=11.55\%$ along the $T_{conf}$ isomorph (Fig. 2 and 3).}
   \label{tableIsomorphPoly12}
  \end{center}
\end{table}

\begin{table}[H]
  \begin{center}
      \begin{tabular}{|l|c|r|}\hline
              $\rho$        &    $D$           &    $\tau$     \\ \hline
              $0.9050$      &    $1100.0000$   &    $10.0000$  \\ \hline
              $1.1000$      &    $3202.8909$   &    $3.0155$   \\ \hline
              $1.3000$      &    $7484.4043$   &    $1.1544$   \\ \hline
              $1.5000$      &    $14801.2586$  &    $0.5306$   \\ \hline
              $1.8000$      &    $33821.6367$  &    $0.2056$   \\ \hline
     \end{tabular}
   \caption{Values of $\rho$, $D$ and $\tau$  for polydispersity $\delta=23.09\%$ along the $T_{conf}$ isomorph (Fig. 2 and 3).}
   \label{tableIsomorphPoly23}
  \end{center}
\end{table}

\begin{table}[H]
  \begin{center}
      \begin{tabular}{|l|c|r|}\hline
              $\rho$        &    $D$           &    $\tau$     \\ \hline
              $0.8500$      &    $1100.0000$   &    $10.0000$  \\ \hline
              $1.0000$      &    $2704.4638$   &    $3.6497$   \\ \hline
              $1.2000$      &    $6991.3804$   &    $1.2502$   \\ \hline
              $1.4000$      &    $14743.5375$  &    $0.5350$   \\ \hline
              $1.7000$      &    $35809.7818$  &    $0.1935$   \\ \hline
     \end{tabular}
   \caption{Values of $\rho$, $D$ and $\tau$  for polydispersity $\delta=28.87\%$ along the $T_{conf}$ isomorph (Fig. 2 and 3).}
   \label{tableIsomorphPoly29}
  \end{center}
\end{table}

\begin{table}[H]
  \begin{center}
      \begin{tabular}{|l|c|r|}\hline
              $\rho$        &    $D$           &    $\tau$     \\ \hline
              $0.9900$      &    $1100.0000$   &    $10.0000$  \\ \hline
              $1.2000$      &    $2798.8895$   &    $3.4571$   \\ \hline
              $1.4000$      &    $5629.7965$   &    $1.5509$   \\ \hline
              $1.7000$      &    $13101.0610$  &    $0.5855$   \\ \hline
              $2.0000$      &    $26062.3855$  &    $0.2641$   \\ \hline
     \end{tabular}
   \caption{Values of $\rho$, $D$ and $\tau$  for polydispersity  $\delta=11.55\%$ along the DIC isomorph (Fig. 5 and 6).}
   \label{tableDICIsomorphPoly12}
  \end{center}
\end{table}

\begin{table}[H]
  \begin{center}
      \begin{tabular}{|l|c|r|}\hline
              $\rho$        &    $D$           &    $\tau$     \\ \hline
              $0.9050$      &    $1100.0000$   &    $10.0000$  \\ \hline
              $1.1000$      &    $2802.7296$   &    $3.4460$   \\ \hline
              $1.3000$      &    $5933.8578$   &    $1.4561$   \\ \hline
              $1.5000$      &    $11036.5957$  &    $0.7116$   \\ \hline
              $1.8000$      &    $23877.3029$  &    $0.2913$   \\ \hline
     \end{tabular}
   \caption{Values of $\rho$, $D$ and $\tau$  for polydispersity $\delta=23.09\%$ along the DIC isomorph (Fig. 5 and 6).}
   \label{tableDICIsomorphPoly23}
  \end{center}
\end{table}

\begin{table}[H]
  \begin{center}
      \begin{tabular}{|l|c|r|}\hline
              $\rho$        &    $D$           &    $\tau$     \\ \hline
              $0.8500$      &    $1100.0000$   &    $10.0000$  \\ \hline
              $1.0000$      &    $2377.8887$   &    $4.1509$   \\ \hline
              $1.2000$      &    $5377.2401$   &    $1.6255$   \\ \hline
              $1.4000$      &    $10460.7186$  &    $0.7540$   \\ \hline
              $1.7000$      &    $23709.0934$  &    $0.2923$   \\ \hline
     \end{tabular}
   \caption{Values of $\rho$, $D$ and $\tau$ for polydispersity $\delta=28.87\%$ along the DIC isomorph (Fig. 5 and 6).}
   \label{tableDICIsomorphPoly29}
  \end{center}
\end{table}

\end{document}